\documentclass[conference]{IEEEtran}
\IEEEoverridecommandlockouts
\usepackage{cite}
\usepackage{hyperref}
\usepackage{amsmath,amssymb,amsfonts}
\usepackage{algorithmic}
\usepackage{graphicx}
\usepackage{textcomp}
\usepackage{xcolor}
\usepackage[table]{xcolor}
\def\BibTeX{{\rm B\kern-.05em{\sc i\kern-.025em b}\kern-.08em
    T\kern-.1667em\lower.7ex\hbox{E}\kern-.125emX}}
\begin{document}

\makeatletter
\def\ps@IEEEtitlepagestyle{%
\def\@oddfoot{\mycopyrightnotice}%
\def\@evenfoot{}%
}
\def\mycopyrightnotice{%
{\footnotesize 978-8-3315-4419-5/25/\$31.00~\copyright2025 IEEE \hfill} 
\gdef\mycopyrightnotice{}
}

\begin{titlepage}

    \thispagestyle{empty} 

    \vfill 

    \centering 
    
    This paper is a preprint; it has been accepted for publication in 2025 IEEE International Conference in Electronic Engineering \& Information Technology, 04-06 June 2025, Chania, Greece
    \par
    \vspace{1cm} 
    IEEE copyright notice © 2025 IEEE
    \par
    \vspace{0.5cm}

    Personal use of this material is permitted. Permission from IEEE must be obtained for all other uses, in any current or future media, including reprinting/republishing this material for advertising or promotional purposes, creating new collective works, for resale or redistribution to servers or lists, or reuse of any copyrighted component of this work in other works.

    \vfill 

\end{titlepage}

\title{Leveraging Digital Twin-as-a-Service Towards Continuous and Automated Cybersecurity Certification \\


\thanks{This work has received funding from the EU Horizon Europe Programme COBALT under Grant Agreement No.101119602}

}

\author{\IEEEauthorblockN{Ioannis Koufos\IEEEauthorrefmark{1},
Abdul Rehman Qureshi\IEEEauthorrefmark{2},
Adrian Asensio\IEEEauthorrefmark{2},
Allen Abishek\IEEEauthorrefmark{3},
Efstathios Zaragkas\IEEEauthorrefmark{1},\\
Ricard Vilalta\IEEEauthorrefmark{3}, 
Maria Souvalioti\IEEEauthorrefmark{1},
George Xilouris\IEEEauthorrefmark{1}, and 
Michael-Alexandros Kourtis\IEEEauthorrefmark{1}
}

\IEEEauthorblockA{\IEEEauthorrefmark{1}Institute of Informatics and Telecommunications, National Centre for Scientific Research ”Demokritos”, Agia Paraskevi, Greece\\
\IEEEauthorrefmark{2}Advanced Network Architectures Lab, Universitat Politècnica de Catalunya, CRAAX-UPC, Spain.\\
\IEEEauthorrefmark{3}Centre Tecnologic de Telecomunicacions de Catalunya - CERCA (CTTC-CERCA), Casteldefells, Spain}}

\maketitle

\begin{abstract}

Traditional risk assessments rely on manual audits and system scans, often causing operational disruptions and leaving security gaps. To address these challenges, this work presents Security Digital Twin-as-a-Service (SDT-aaS), a novel approach that leverages Digital Twin (DT) technology for automated, non-intrusive security compliance. SDT-aaS enables real-time security assessments by mirroring real-world assets, collecting compliance artifacts, and creating machine-readable evidence. The proposed work is a scalable and interoperable solution that supports open standards like CycloneDX and Web of Things (WoT), facilitating seamless integration and efficient compliance management. Empirical results from a moderate-scale infrastructure use case demonstrate its feasibility and performance, paving the way for efficient, on-demand cybersecurity governance with minimal operational impact.

\end{abstract}


\section{Introduction}

The cyber threat landscape has become increasingly complex and is marked by the integration of new pervasive technologies, which have introduced significant security challenges. Industry professionals underscore the urgent need to adopt robust cybersecurity strategies and practices to navigate through such a continuously evolving threat environment \cite{google2025cybersecurity}. A recent research from the Ponemon Institute outlines that 52\% of their respondents believe that the most important governance activity is to carry out frequent internal or external audits for cybersecurity and information security compliance \cite{optiv2024cybersecurity}. While such a need has been previously emphasized to be a good practice, nowadays it has become a necessity with the NIS-2 directive.  NIS-2 mandates rigorous audit practices as a core component of cybersecurity governance, aiming to increase overall cyber resilience across the European Union.

Traditional risk assessments rely heavily on manual processes, which may require extensive system scans, log collection, and manual configuration reviews. Such processes consume system resources, slow down operations, and even introduce downtime if they require stopping or isolating certain components for assessment. Moreover, traditional risk management follows a periodic assessment model, meaning that vulnerabilities might go undetected between audits, while conducting on-demand or large-scale security assessment can put a strain on the Information Technology (IT) infrastructure, potentially affecting availability \cite{chidukwani_2022}.

In recent years, the use of Digital Twins (DT) as a solution that facilitates decision-making and emulation, without impacting real-world operation, has been attracting the interest of the research community. A DT is a virtual model of physical or cyber assets and processes, synchronized at a certain time frequency with respect to the audit needs. According to the DT Consortium's architecture \cite{heaton_platform_nodate}, the virtual model can contain both static and dynamic information, represented as stored and computational data, respectively. Initially, DTs were widely used in aerospace, manufacturing, and telecommunications for testing, verification, and validation before real-world implementation. The advent of Industry 4.0 (I4.0), integrating technologies such as AI, Big Data, Cloud, and IoT, is driving the rapid advancement of DT technology across various domains. Interestingly, a particularly noteworthy domain where the use of DTs is recently attracting the interest of researchers is cybersecurity \cite{jaber_2025}.

\begin{figure}[h]
    \centering
    \includegraphics[width=0.8\columnwidth]{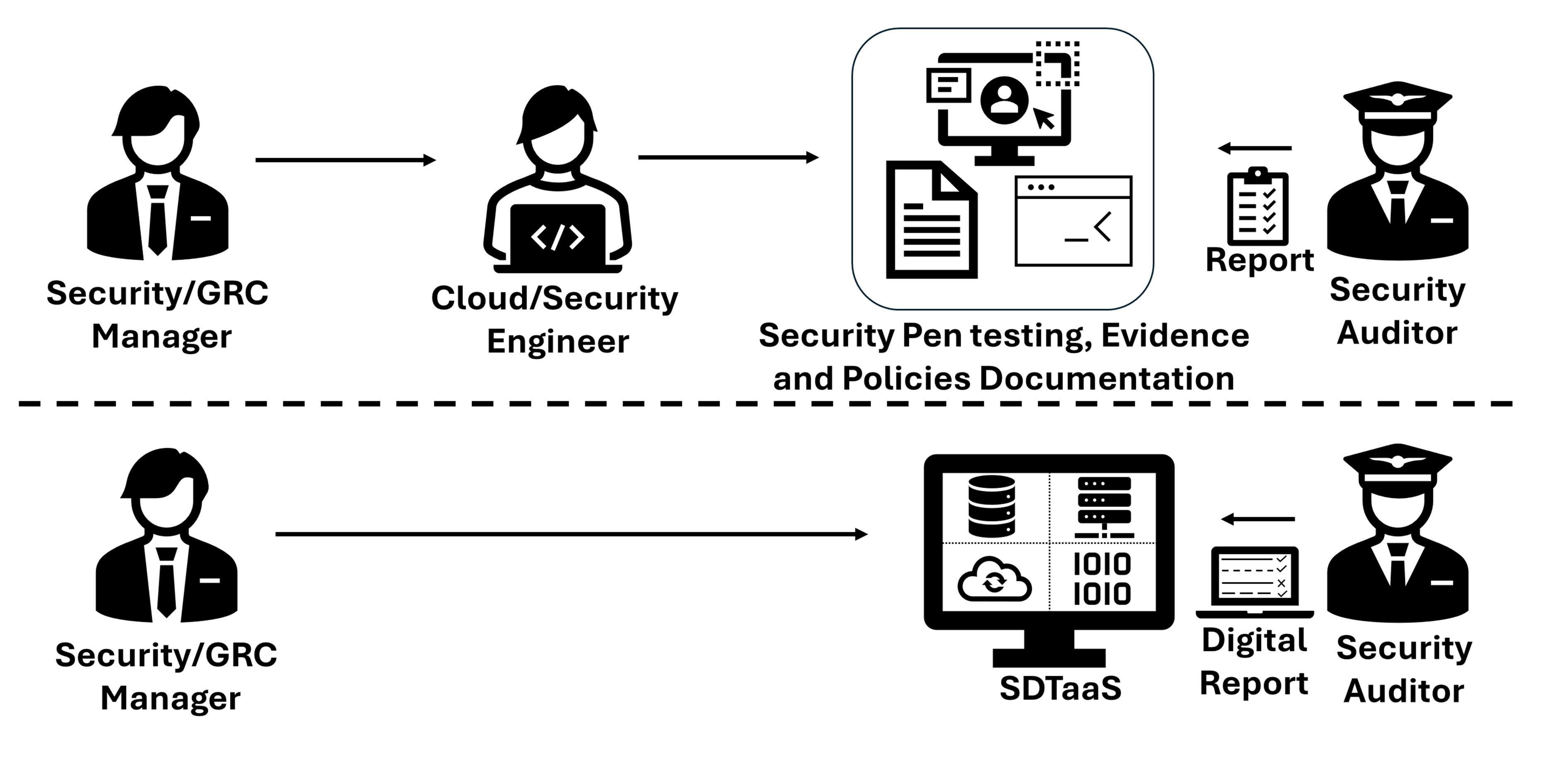} 
    \caption{An illustration of SDT-aaS benefits\cite{zech2024digital}.}
    \label{fig:dtaas-benefits}
\end{figure}

Based on the DT concept, the work presented in this paper is a structured approach to automation, abstraction, and emulation in information security audit through a new innovative concept, called Security Digital Twin-as-a-Service (SDT-aaS). As it is represented in Figure~\ref{fig:dtaas-benefits}, the proposed solution aims to automate the deployment of services that mirror assets from a computer network, while reducing the technical expertise required by governance, risk, and compliance security managers to assess system security through digital reports. 

Additional advantages include customization for expanded capabilities and improved interoperability with open standards such as CycloneDX, and Web of Things for the virtual representation of assets. These features constitute an on-demand, scalable solution for security assessments in complex system architectures. The powerful combination of CaC support, automated risk management, and SDTs aims to enhance system availability by enabling continuous compliance monitoring at short, periodic intervals. This approach ensures non-intrusive security checks, minimizing disruptions to operations while avoiding the need for full system scans. The main contributions of this work are:
\begin{itemize}
    \item A novel architecture for non-intrusive and automated evidence collection delivery of SDT-aaS for continuous information risk management.
    \item The development of an adaptable, fast, and scalable DT-aaS management system that can operate for diverse infrastructures.
    \item The experimental validation of the proposed architecture in a realistic small-medium business (SMB) digital infrastructure.
\end{itemize}

\section{Related Work}

Throughout the years, many attempts have been made to tackle the known issue of compliance in complex ICT environments. The core difficulty stems from the need to integrate a vast array of well-defined requirements, as outlined in established risk management frameworks such as ISO/IEC 27001 and NIST 800-37. As a result, many recent works have focused on automating distinct risk management processes, placing particular emphasis on approaches that do not harm the target's system availability. In particular, the authors in \cite{gill_2021} examine NIST 800-53 Controls and conduct a detailed analysis of which security and privacy controls can be fully automated. Although 47.9\% of high-impact security and privacy controls can be automated, the authors explain that human intervention is mandatory and that the feasibility of risk management automation can be established through strict criteria (e.g., machine-readable asset definitions).
Similarly, the authors in \cite{sterbak_2021} highlight the need of automated asset discovery tools to replace manual identification and dependency analysis in complex environments. Furthermore, they emphasize that such automation must be accompanied by continuous monitoring and control mechanisms to effectively manage risks and assets in real time.

A promising approach that addresses these challenges is the concept of Compliance-as-Code, CaC, which aims to embed compliance checks throughout the system's operational lifecycle \cite{agarwal_2022}. This is achieved by replacing the manual identification of assets and corresponding risks with machine-readable artifacts. Machine-readable artifacts are automatically created for identified environmental assets and can be associated with established security and privacy control deficiencies and Efficiencies against a target system.

Recent advances in automated compliance \cite{masi2023} showcased the integration of DT technologies for automated compliance validation and real-time monitoring. The proposed system in that work is able to compare operational data against compliance criteria, thus reducing significantly the workload of manual audits. 

Aligned with compliance and risk management framework requirements, DTs can be used for cyber threat intelligence, threat detection, simulation, testing, automation, and security-based asset management. The authors of \cite{ dietz_harnessing_2022,dietz_employing_2022} established the foundation for using DTs to detect potential threats in cyber-physical systems and industrial control systems. However, the proposed approach does not provide a comprehensive security posture of the system. Interestingly, the works in \cite{empl_soar4iot_2022,empl_digital-twin-based_2023}, focus on modeling IoT devices as DTs. These DTs are then synchronized and mirrored in a centralized asset management platform, Eclipse Ditto\footnote{\url{https://eclipse.dev/ditto/}}. This platform can act as middleware to integrate various security applications to reduce effort, provide a complete security posture of the system, and facilitate further analysis for compliance.

However, to avoid incurring ever-growing costs due to the continuous mirroring of the real system in the virtual domain, cost-efficient scalable solutions need to be explored. To that end, solutions offering DT-as-a-Service (DT-aaS) are found in the literature, facilitating on-demand DT creation while enabling real-time monitoring, lifecycle management, and fine-grained control of assets through cloud-based platforms \cite{aheleroff_digital_2021}. Initially explored for I4.0, the concept of DT-aaS aims at enhancing DT creation, management, and deployment, offering a scalable and efficient approach \cite{zech2024digital}. Furthermore, major platforms, such as AWS TwinMaker\footnote{\url{https://aws.amazon.com/iot-twinmaker/}}, Microsoft Azure DT\footnote{\url{https://azure.microsoft.com/en-us/products/digital-twins/}} for IoTs DTs, and Cybellum\footnote{\url{https://cybellum.com/cyber-digital-twins/}} for vulnerability management for product security through DTs. Nevertheless, current solutions often lack full interoperability and customization\cite{zech2024digital}.

Beyond virtual asset availability, seamless DT deployment is key to reducing time, effort, and cost. Aligned with DT-based architectures that can be found in the literature (e.g., \cite{kherbache_digital_2022}), and with the concept of DT-aaS, we present an SDTaaS architecture to facilitate compliance, by automatically gathering and integrating Bill of Materials (BOMs) from software assets from ICT infrastructures into a DT. To that end, we introduce the concept of Security DT (SDT), acting as a single source of truth and mainly exposing BOMs, as CaC artifacts, mirrored from assets from the ICT infrastructure, to audit and compliance services executed on top of it. Next, the section presents our proposed architecture.

\section{Proposed Architecture}
In this section, an overview of the proposed architecture, supporting the automation of relevant audit management services, and a detailed description of its main components and their functionalities are provided.

\begin{figure*}[t]
\centerline{\includegraphics[scale=0.44]{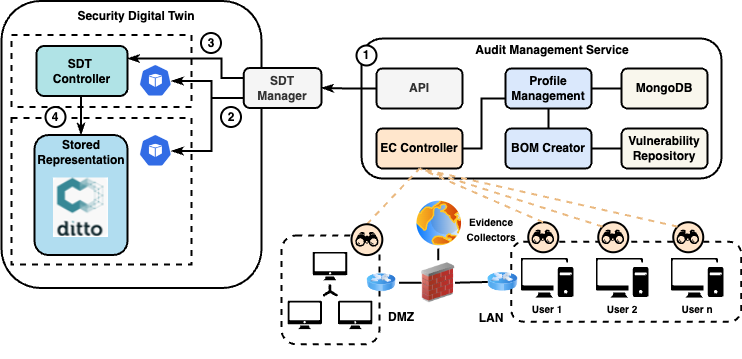}}
\centering
\caption{High-level architecture representing the real system, the Audit Management Service, and the components to manage SDT-aaS.}
\label{fig:arch}
\end{figure*}

\subsection{Architecture Overview}
The proposed architecture to facilitate non-intrusive compliance based on SDT-aaS is illustrated in Figure \ref{fig:arch}, in which three main domains can be identified: i) the real system, that is a computer network,  ii) the services to support audit management automation, and iii) the DT domain, supporting the SDT-aaS. In addition to the components depicted in the figure, the specific services around security compliance need to be deployed on top of the SDT instances.

We assume a heterogeneous computer network, composed of a wide variety of hardware, software, and network assets, implementing different services. On top of it, the Audit Management System (AMS), is responsible for automating the deployment of agents that collect sufficient evidence from the different devices in the computer network. Specifically, these agents target gathering relevant information from the assets aimed at building their BOMs, and recreate the underlying topology, from relationships identified among the different assets. We highlight the relevance of this component in automating and abstracting the complexity of building the BOMs of the computer network and recreating its topology. Regarding the DT domain, we propose an SDT Manager that handles requests from the AMS to deploy SDT instances, on demand, aimed at supporting compliance services without impacting the normal operation of the underlying computer network. Next, the AMS and the components to support the SDT-aaS are detailed.

\subsection{Audit Management Service}
The Audit Management Service, AMS, component is comprised of 3 core sub-components (Figure \ref{fig:arch}): i) the Profile Management System, ii) the BOM Creator, and iii) the Evidence Collector (EC) Controller and servers three main purposes.

The first purpose is to define and select an audit profile and conduct asset identification, SDT instantiation, and event-driven SDT updates. Specifically, the Profile Management service starts by identifying target hosts and mapping the topology to a non-relational database (e.g., MongoDB). It also creates a profile for each host and its corresponding digital assets of interest, in accordance with the selected audit profile.

The second purpose is to create and orchestrate evidence collectors to capture important information such as risks and threats that need to be treated. The EC Controller deploys agents to the hosts belonging to the selected audit profile. These agents search the target system and gather meaningful audit artifacts and their underlying inter-dependencies. Examples of audit artifacts are summarized in Table \ref{tab:crypto_audit}.

\begin{table*}[htbp]
    \centering
    \caption{Example Audit Artifacts Gathered by the Evidence Collectors}
 \begin{tabular}{|p{4cm}|p{6cm}|p{6cm}|}
        \hline
        \textbf{Category} & \textbf{Description} & \textbf{Indicative Examples} \\
        \hline
        Cryptographic Libraries & List of cryptographic libraries supported by the system & OpenSSL, BouncyCastle, Crypto++, mbedTLS \\
        \hline
        Digital Certificates & Digital certificates stored and used for authentication & SSL/TLS certificates, X.509 certificates \\
        \hline
        Utilized Algorithms & Cryptographic algorithms utilized by the system & AES-256, RSA-4096, ECC, SHA-256 \\
        \hline
        OpenSSL & OpenSSL configurations and supported protocols & Enabled TLS versions, Cipher Suite settings \\
        \hline
        Kernel Settings & Kernel-level security and cryptographic configurations & sysctl settings for entropy pool size, FIPS mode \\
        \hline
        System Logs & System logs capturing cryptographic and access control related events & Authentication logs, encryption failure logs \\
        \hline
        Software Projects supporting various programming languages, project and library dependencies & Java, PHP, Python, and JavaScript projects with configurations, plus external libraries and dependencies & Spring Boot, Hibernate, Apache Kafka, Laravel, Flask, Django, FastAPI, Express.js, React, Angular, Vue.js, Maven, PHP Composer, pip, NPM/Yarn   \\
        \hline
    \end{tabular}
    \label{tab:crypto_audit}
\end{table*}

Finally, the third purpose of the AMS is to translate the aforementioned information into a lightweight and standardized machine-readable document that facilitates seamless integration and automation of audit procedures. That is, the proposed AMS processes audit artifacts to create BOMs for the identified assets. In our approach, BOMs follow the OWASP CycloneDx\footnote{https://cyclonedx.org} v1.6 standard format (currently an ECMA-424-International Standard\footnote{https://ecma-international.org}) and support Software BOMs (SBOMs), Cryptography BOMs (CBOMs), Vulnerability Exploitability Exchange (VEX) and BOM-Links to describe connections, relations, and dependencies. Focusing on CBOMs, the BOM creator employs a hierarchical graph representation of cryptographic components. Therefore, information is well-structured and consistent, thus forming an interconnected system where algorithms, protocols, and key attributes are systematically organized and enriched with each newly discovered entry. This hierarchical model enhances the ability to analyze cryptographic relationships, ensuring that dependencies are well-defined and navigable, which is especially relevant when a CBOM captures a large volume of information. Consequently, the AMS automatically connects BOMs with the target audit profile. Moreover, to fill vulnerability-related attributes of VEX files, this component uses an internal vulnerability database, storing Common Vulnerability Exposure\footnote{https://cve.mitre.org} (CVE) information and the corresponding Common Vulnerability Scoring System\footnote{https://www.first.org/cvss/} (CVSS) metrics, for already identified vulnerabilities. 

\subsection{SDT Manager and SDT-aaS}
Regarding the DT domain, the two main components are the SDT Manager and the SDT instances. Specifically, the SDT Manager is responsible for decentralized SDT creation, guided by strategies that optimize task allocation throughout the SDT lifecycle. In addition, the SDT Manager coordinates with the AMS to accommodate any updates that emerge from re-scoping the target of evaluation or from changes on the real-world counterpart, ensuring that all relevant information is accurately mirrored on the SDT.

The primary input to SDT Manager comprises an HTTP request (structured in JSON) received from the AMS, which contains the core data necessary for SDT initialization or updates. In response, the SDT Manager relays pertinent information back to the AMS, confirming the successful handling of requests or signaling the need for additional adjustments. Furthermore, the SDT Manager issues HTTP requests to individual SDTs both for disseminating data received from the AMS for state synchronization, and for managing SDT operations and lifecycle activities. All interactions are facilitated through a robust HTTP API, ensuring secure and reliable communication among components.

The internal architecture of the SDT Manager includes the SDT Manager Interface, the SDT Manager Core, the SDT LifeCycle Manager (LCM), and a Data Adapter.

The SDT Manager Interface operates as a secure communication channel, enabling the reception of requests and data from the AMS and the transmission of the associated responses. At the core of this architecture lies the SDT Manager Core, which orchestrates workflows such as SDT creation, update processes, and overarching coordination with other subsystems. The SDT LCM is dedicated to configuring and deploying SDT instances, incorporating tasks such as resource allocation. 
Finally, the Data Adapter component processes the data channeled to the SDT, preserving consistent and up-to-date representation within each SDT instance. By integrating these components, the SDT Manager ensures that digital replicas remain accurate, secure, and responsive to the dynamics of the underlying assets and corresponding processes being represented.

Regarding the SDT instance, we consider an approach built on two main components: i) the SDT Controller, implementing the integration representation/function and a service interface to facilitate the interaction with the SDT from external services through a robust access control policy, and ii) the stored representation, representing the states and attributes of the computer network assets along the time.

\begin{figure*}[t]
\centerline{\includegraphics[scale=0.35]{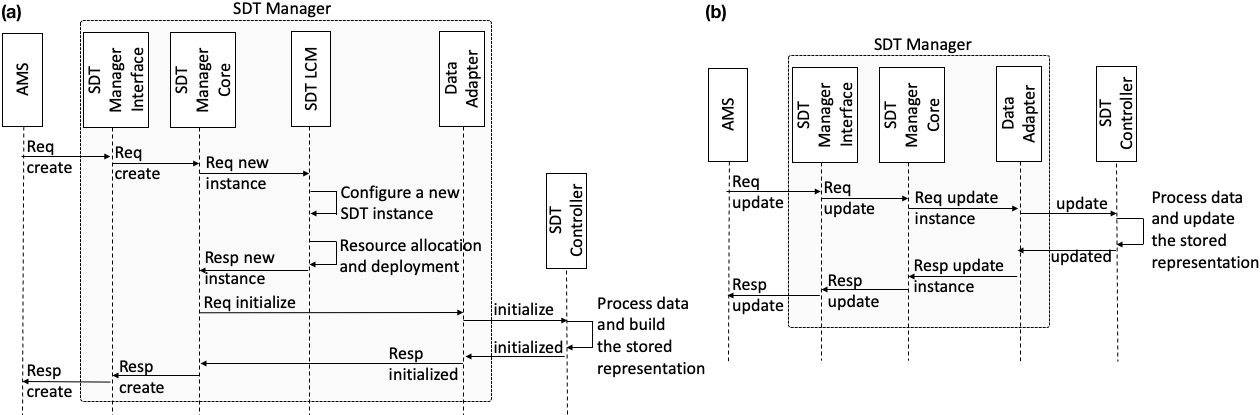}}
\centering
\caption{Sequence Diagram with SDT Manager internal components for SDT creation (a) and update (b).}
\label{fig:seq-sdt}
\end{figure*}

\subsection{Sequence Diagram}

Figure \ref{fig:seq-sdt}a depicts the sequence of interactions involved in the creation and deployment of an SDT instance. The diagram begins with the AMS initiating the process by sending a request to the Interface in the SDT Manager (step numbered as 1, in Figure \ref{fig:arch}), which requests the SDT Manager Core to trigger the creation of an SDT using the received data from the AMS. The SDT Manager Core then interacts with the SDT LCM to configure and deploy a new SDT instance. The SDT LCM proceeds by deploying the SDT (step 2, in Figure \ref{fig:arch}).
Following this, the SDT Manager Core instructs the Data Adapter component to process the data, i.e., SBOMs and CBOMs, and to build the stored representation through the SDT Controller (steps 3 and 4 in Figure \ref{fig:arch}). Once the stored representation is successfully built, the Data Adapter confirms this back to the SDT Manager Core, which sends a response back to the AMS through the Interface, indicating that the SDT has been successfully created. Figure \ref{fig:seq-sdt}b illustrates the sequence of interactions involved in updating the stored representation of an SDT instance from the SDT Manager. The process begins with the AMS sending a request to the SDT Manager Interface, which delegates it to the SDT Manager Core to update the SDT with new data, representing the current state of the assets in the computer network. The SDT Manager Core then forwards this request to the Data Adapter component to update the corresponding SDT instance, through its SDT Controller.

Next, we validate the proposed architecture, implementing it in an experimental environment and considering Eclipse Ditto, which supports the Web of Things\footnote{https://www.w3.org/WoT/} description, to handle the stored representation and Kubernetes as the technology to support the deployment of SDT instances. The AMS, the SDT Manager, and the SDT Controller are implemented in Python.

\section{Validation and Experimental Results}
\label{section: validation and experimental results}
In this section, we outline our approach to evaluate the proposed SDT-aaS within the context of organizations preparing for an internal on-demand audit. Our assessment involves comparing the performance of SDT instantiation and building its stored representation using two different topologies: i) a topology representing an organization managing only essential services and users, and used for validation and as reference for performance comparison, and ii) a more complex topology, representing a realistic scenario reflecting an SMB with multiple services and hosts, used for performance evaluation, under a realistic approach. The VM used in the experiment is provisioned with 8 vCPUs (clocked at 2.75 GHz), 16 GB of RAM and 80 GB of disk and runs Ubuntu Server 22.04.5 LTS.

\subsection{Validation}
\begin{figure}[t]
\centerline{\includegraphics[width=0.9\linewidth]{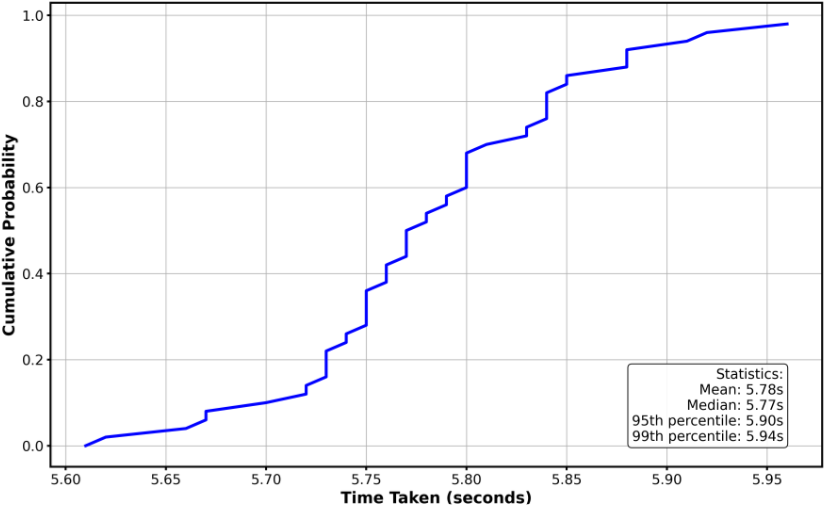}}
\centering
\caption{CDF for the time taken to deploy SDT-as-a-Service}
\label{fig:SDT_deploy}
\end{figure}

For validation purposes, we implemented a virtualized topology, supporting a few users, without relying on complex or specially configured hosts, and including typical services running on a single server machine. To carry out the experiments, SDT instances were deployed via the SDT Manager 50 times. Figure ~\ref{fig:SDT_deploy} shows the Cumulative Distribution Function (CDF) of the time taken to deploy an SDT via an API call from the SDT Manager. A remote client (i.e., the AMS) can send a POST request to the SDT Manager which starts the process of SDT instantiation. The SDT is assigned a unique hexadecimal identifier which can later be used to reference it. The SDT is instantiated as a pod in a Kubernetes cluster.

As it is shown in Figure ~\ref{fig:SDT_deploy}, the mean time to deploy an SDT was 5.78 seconds, while the minimum time was 5.61 seconds and the maximum was 5.96 seconds, which shows consistency and very low variance of mean time compared to each iteration. It must be noted that the SDT pod in the Kubernetes cluster can also be destroyed through an API call from the SDT Manager, facilitating SDT-aaS functionality.

\subsection{Performance Evaluation}
\begin{figure}[t]
\centerline{\includegraphics[width=0.9\linewidth]{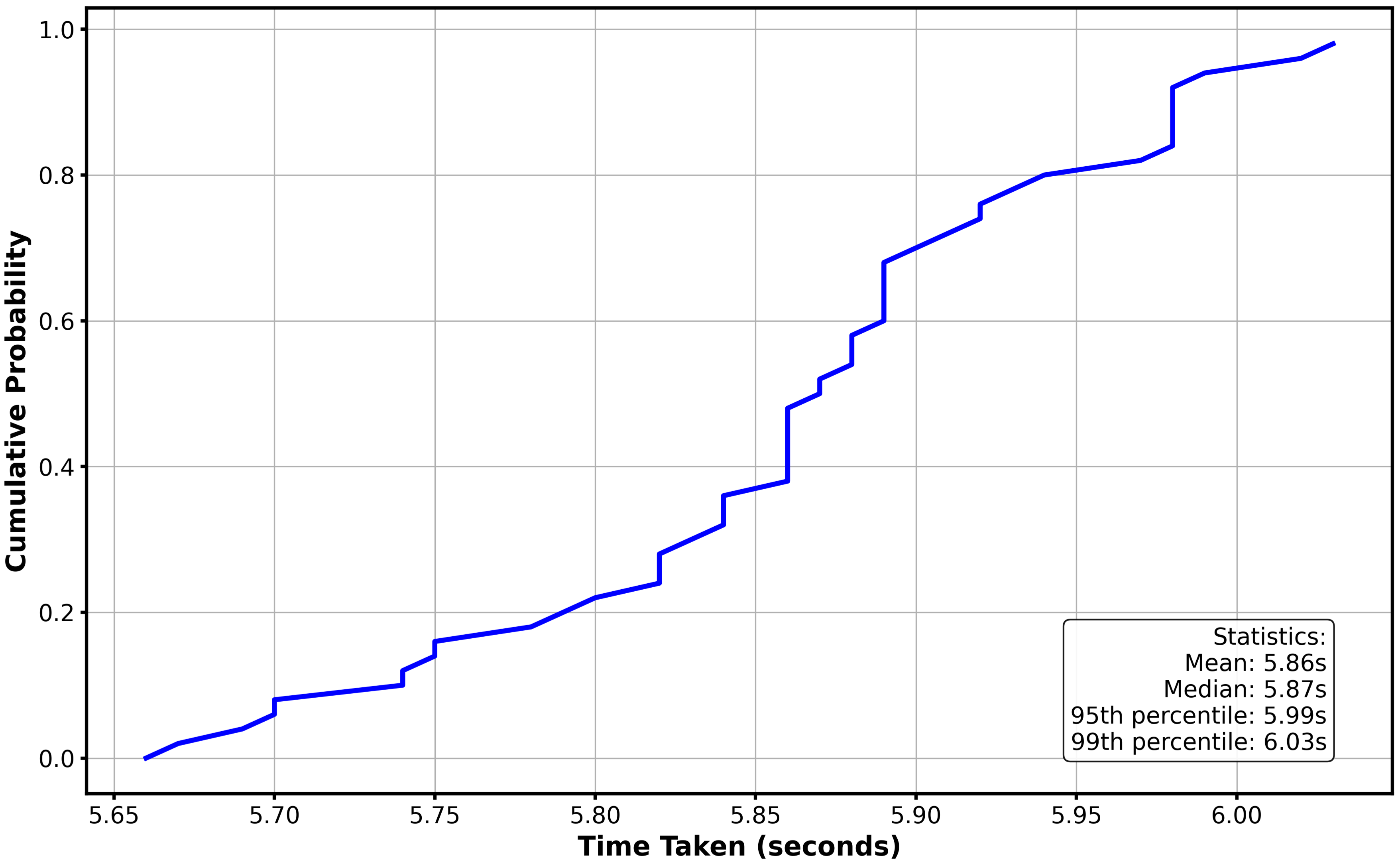}}
\centering
\caption{CDF for time to deploy a larger size SDT}
\label{fig:SDT_demokritos_deploy}
\end{figure}

For completeness, we implemented a setup assuming a DMZ, and hosting a web server, mail servers, the organization's Software-as-a-Service (SaaS) application, multiple Spring Boot Microservices, and a couple of hosts used for development purposes on a separate LAN, as summarized in Table\ref{tab:evidences-collected-topology-medium}. An SDT with the services and components mentioned in Table \ref{tab:evidences-collected-topology-medium} was instantiated as a service on the same Ubuntu server considered for validation. As previously described, an API call (POST) is issued to the SDT Manager with instructions to instantiate an SDT. This triggers the instantiation of the SDT as a pod in a Kubernetes cluster. Figure \ref{fig:SDT_demokritos_deploy} shows the CDF plot of instantiating the SDT with the audit artifacts shown in Table \ref{tab:evidences-collected-topology-medium}.

\begin{table}[htbp]
    \centering
    \caption{Total CaC Artifacts for SMB Topology}
    \begin{tabular}{|p{2.8cm}|p{4.5cm}|}
        \hline
        \textbf{Machine/Host} & \textbf{Audit Artifacts} \\
        \hline
        Web Server & Algorithms (8), Vulnerabilities (24), Components (29), Certificates (1) \\
        \hline
        Spring MicroServices & Algorithms (8), Vulnerabilities (12), Components (9), Certificates (0)\\
        \hline
        Management System & Algorithms (10), Vulnerabilities (7), Components (6), Certificates (0) \\
        \hline
        Mail Server & Algorithms (9), Vulnerabilities (0), Components (4), Certificates (1) \\
        \hline
        Users (Total) & Algorithms (23), Vulnerabilities (153), Components (75), Certificates (3))\\
        \hline
    \end{tabular}
    \label{tab:evidences-collected-topology-medium}
\end{table}

Figure \ref{fig:SDT_demokritos_deploy} was obtained by performing 50 iterations of SDT instantiation. It can be seen that the mean time to instantiate the SDT is 5.86 seconds with a median time of 5.87, such low variance stands proof to be a consistent process of SDT instantiation. Compared to the reference scenario, where the mean time for instantiation of the (lighter) SDT, is 5.78 seconds (Figure \ref{fig:SDT_deploy}), results showed that the vertically scaled SDT takes an additional time of 80 milliseconds to be instantiated. This can be attributed to the extra JSON files that need to be processed, representing a realistic scenario and resulting in a larger SDT. In addition, the total storage footprint used for the SMB topology is measured in just a few MBs for a complete virtual replica of the identified assets and configurations. This reduced size is mostly achieved by the coherent and organized structure of OWASP CycloneDX standard format.

\section{Conclusions \& Future Work}
As the cyber threat landscape is rapidly evolving, there is a great demand for robust security governance and compliance. Traditional cybersecurity audits and risk assessments rely on manual processes which are often resource-intensive, disruptive, and slow. DT-aaS has emerged as a promising solution for information security compliance and risk management.

In this work, we introduced Security Digital Twin-as-a-Service for the automation of on-demand security audits that are adaptable to various infrastructures. SDT-aaS is deployed and instantiated through a proposed Audit Management Service, paving the way to a more efficient, accessible and seamless evidence collection for security professionals by integrating, Compliance-as-Code and existing cybersecurity frameworks (CVE, CVSS).
Our approach is highly scalable, enables rapid and effortless deployment, and at the same time maintains a lightweight footprint with minimal data storage, as outlined in Section \ref{section: validation and experimental results}. Moreover, our solution offers improved interoperability, due to the integration of open standards, such as OWASP CycloneDX and Web of Things.

Future work would explore the incorporation of more open standards for the representation of evidence artifacts in order to provide a more detailed and complete version of the existing work, as the performance of our work is consistent through the usage of CaC. We are also looking forward to investigate continuous resource monitoring, conduct large-scale studies and empirical evaluations to further asses the effectiveness and scalability of SDT-aaS along with foreseen automated compliance verification (e.g, using the Open Security Control Assessment Language)\footnote{https://pages.nist.gov/OSCAL/}.

\bibliographystyle{ieeetr}
\bibliography{bibliography}

\vspace{12pt}

\end{document}